# Function of Forgetfulness for the Tedium of Oblivion on Liquidity of Ontology Matching

Massimiliano Dal Mas
me @ maxdalmas.com

**ABSTRACT**
The shallow and fragile knowledge on the Web does not examine in depth the things: it behaves lightly. The conditions created by the Web makes our attention labile and especially fickle, it's unable to concentrate for long as we are trained to "surf" without going though never in depth.
The Web also brings with it the added advantage of a nearly availability infinite knowledge but leads to a loss of the ability to retain and evaluate that knowledge within us increasing forgetfulness of knowledge.
In this paper we show how the "function of forgetfulness" appears linked to tedium and oblivion of knowledge through the liquidity of ontology matching.

*Keywords:* Ontology; ontology matching; usage mining; similarity; user behavior; dynamical system; mathematical modeling; linear dynamics, oblivion, tedium.

## INTRODUCTION

With the "liquidity" in global knowledge due to the increasing of the Web the old knowledge follow a rapid oblivion due to an increasing on the development of new knowledge.

The aim of this work was to define a "*function of forgetfulness*" on the tags used to define a text in terms of tedium being rapidly uninteresting or boring.

This work was validated testing 20 users and how they associated tags related to different ontologies during a period of time.

---

If you have read this paper and wish to be included in a mailing list (or other means of communication) that I maintain on the subject, than send e-mail to:
me @ maxdalmas.com

A description of the mathematical model proposed is followed by the experiment done. Finally, some conclusions and future works are introduced.

## THE BASIC IDEA

In this section we see the Basic Idea on the approach followed.

We want to consider the moments when the User changes his interest on a Web Object related to a different Ontology. [1-5] For our work we considered groups of hashtags of different Ontologies defining a Web Object. [6, 7] We look for the time between a different combination of a tag to the same Web Object made by the same User.

## THE MATHEMATICAL MODEL

In the following paragraphs it's explained in detail the processes mentioned above and it is introduced the "*function of forgetfulness*" by which a Web User forget the old knowledge matching the tag chosen by defined ontologies.

### Knowledge in the Web

Internet brings all of the advantages, including a nearly availability infinite knowledge. But the conditions created by Internet makes our attention labile and especially fickle, unable to concentrate that long, coached some 'surfing' without going though never in depth. [9]

In fact with the advent of the Web the messages become short and simple, so as to communicate all their contents before attention runs out. [10] By long, elaborate and thoughtful letters it has gone up to a short email messages even laced up to microblogging with 120-140 characters for "tweet" to Twitter. [11] Where short messages from Twitter can be labeled with the use of one or more "hashtag" - words or combinations of words concatenated preceded by the pound sign (#).

### The Tedium of Oblivion

As stated by Zygmunt Bauman: "The development of culture is based on the learning of new precepts and oblivion of the old ones" [9]. As example imagine if we have to light a fire only with straw and a flint: we do not know how to do it because we do not used, or thought about it any more.

When we buy an item in a store there is an implicit understanding for which we'll substitute it when we are no more satisfied on it. This consumerist model of customer relationship good consumption is increasingly present in relations between human and 'knowledge'. [9] The old knowledge follow an oblivion due to the quality or state of being uninteresting or boring. [12, 13] The oblivion of the knowledge is so correlated to a process of tedium.

### Forgetfulness function

On Internet the *variation of interest xi* of the Web User *i* for the Web Object belonging to an ontology *j* can be correlated to the tedium for that web object. [3, 4, 6, 7] The process of *tedium Mi* can be easily studied considering the extreme case of a Web User who overcome an interaction with the Web Object been neither completed, assuming that in this condition *xi (t) = 0* vanishes exponentially with time (1) is obtained the (2).

$$(1) \quad x_i(t) = x_i(0) e^{-m_i t}$$

$$(2) \quad \dot{x}_i = -m_i x_i$$

$$(3) \quad M_i(x_i) = -m_i x_i$$

Therefore, the "*function of forgetfulness*" appears in (3) where the coefficient of "*tedium*" (*mi*) can only have positive values.

The "*tedium factor*" (*mi*) represents the velocity with which the Web Object is forgotten by the Web User – more *mi* is small and more the velocity is high. From the mathematical point of view it can be

thought as a phenomenon of a linear decreasing (3). [1, 8]

## EXPERIMENT

To validate the assumptions made we proposed to a group of 20 users to associate different tags (related to different ontologies) to the same text during five months. During the interval of time it was observed the changes in the correlation on the tags with the same proposed text. The analysis of the data showed a decrease in the use of tags during the observation time according to (3). In Figure 1 is depicted the trend of the "*forgetfulness function*" showing the linear decreasing phenomenon.

**Figure 1**   Typical trend of Forgetfulness function

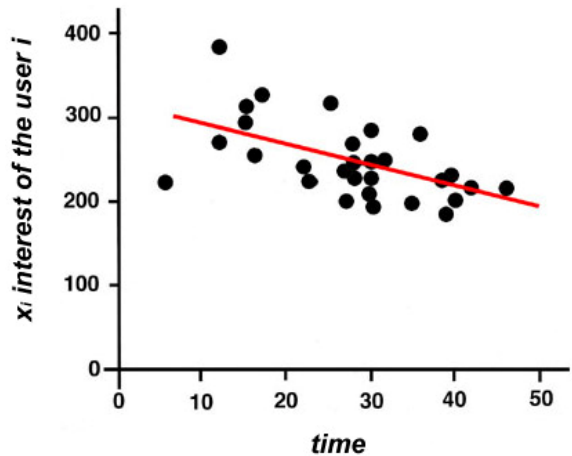

## CONCLUSION

Mathematical modeling inevitably tends to simplify reality but the "function of forgetfulness" represents a first step for the development of a Dynamic Model on Web User Feeling to be used in an Adaptive Ontology Matching [1] for the "liquidity" of the knowledge on the Web [9].

(etails.jsp?arnumber=6245653) DOI: 10.1109/CISIS.2012.158